\documentclass[10pt]{article}

\usepackage{graphicx}
\usepackage{amssymb}
\usepackage{amsmath}
\usepackage{cite}
\usepackage{color} 

\newcommand{\ew}[1]{\langle#1\rangle}
\newcommand{\km}{\ensuremath{\ew{k}}} 

\newcommand{\m}[1]{\mathrm{#1}}
\newcommand{\per}{\mathrm{periphery}}
\newcommand{\co}{\mathrm{core}}
\newcommand{\secl}{\ensuremath{\mathrm{secluded}}}

\begin{document}

\begin{flushleft}
{\Large
\textbf{Limits and trade-offs of topological network robustness}
}
\\
Christopher Priester$^{1,\ast}$, 
Sebastian Schmitt$^{2}$, 
Tiago P.\ Peixoto$^{3}$
\\
\bf{1} Institut f\"ur Festk\"orperphysik, Technische Universit\"at Darmstadt, Hochschulstra\ss e 6-8, D-64289 Darmstadt, Germany
\\
\bf{2} Honda Research Institute Europe GmbH, Carl-Legien-Stra\ss e 30, D-63073 Offenbach/Main, Germany
\\
\bf{3} Institut f\"{u}r Theoretische Physik, Universit\"{a}t Bremen, Hochschulring 18, D-28359 Bremen, Germany
\\
$\ast$ E-mail: christopher@fkp.tu-darmstadt.de

\end{flushleft}

\begin{abstract}
  We investigate the trade-off between the robustness against random and
  targeted removal of nodes from a network. To this end we utilize the
  stochastic block model to study ensembles of infinitely large networks
  with arbitrary large-scale structures.  We present results from
  numerical two-objective optimization simulations for networks with
  various fixed mean degree and number of blocks. The results provide
  strong evidence that three different blocks are sufficient to realize
  the best trade-off between the two measures of robustness, i.e.\ to
  obtain the complete front of Pareto-optimal networks.  For all values
  of the mean degree, a characteristic three block structure emerges
  over large parts of the Pareto-optimal front. This structure can be
  often characterized as a core-periphery structure, composed of a group
  of core nodes with high degree connected among themselves and to a
  periphery of low-degree nodes, in addition to a third group of nodes
  which is disconnected from the periphery, and weakly connected to the
  core.  Only at both extremes of the Pareto-optimal front,
  corresponding to maximal robustness against random and targeted
  node removal, a two-block core-periphery structure or a one-block
  fully random network are found, respectively.
\end{abstract}


\section{Introduction}
\label{sec:intro}

The theoretical investigation of complex networks has proven to be a
valuable tool for the study of many real-world
systems~\cite{albertNetworks02,allesinaStability12,
boccalettiNetworks06,carvalhoRobustGas09,
  corneliusNetworkControl13, albertPower03}.
One important aspect is how the topological properties of networks are
linked to their function and robustness~\cite{callaway_network_2000,
buldyrevCascades10}. Robustness is defined as the correct functioning in
the presence of disturbances, and it is a desired property of many
empirical network systems. The robustness of networks to topological
disturbances is a very active field of
research\cite{casalsTopologyPowerNet07,buldyrevCascades10,
  hinesTopology10,schneiderAttack11}, since it is often assumed that it
is a necessary ingredient for higher-order forms of robustness
associated with specific network dynamics \cite{SzejkaRobust09,
LiYeast04, ScottRobustness06, DoddsOrganization03, peixoto_emergence_2012}.

One popular way to address topological robustness is by removing nodes
from a given network and then analyzing how connected the network
remains as function of the number of nodes
removed~\cite{cohen_resilience_2000,callaway_network_2000,cohen_breakdown_2001}.
In this way, the problem of robustness is mapped to the classical
phenomenon of percolation, and the formation of a giant component in the
remaining network after the node removals.

Recent studies focused on the optimization of the topological robustness
of networks, when a given set of constraints are imposed
\cite{mathiasSmallWorldOpt01,canchoOptimizationNetwork03,colizzaSelection04,
  netoteaEvoRobust06,bredeOpt10,bredeNetworkOpt09,schneiderAttack11,peixotoRobustTopology12}.
Most recent works have focused on optimization according to different
robustness criteria, such as targeted
attacks~\cite{ValentePeak04,schneiderAttack11,peixotoRobustTopology12}
and random
failure~\cite{ValentePeak04,peixotoRobustTopology12}. However, most real
systems are subject to simultaneous types of perturbations, which
individually require different, and thus competing strategies to
mitigate failure. In order to properly access the inherent trade-offs in
such situations, one needs to combine multiple robustness criteria. A
standard technique is to chose a weighted sum of the relevant criteria
as the objective function to be minimized or maximized. However, such an
approach can be ineffective if the goal is to map all possible trade-off
values between these objectives.  In addition, it also bears the
difficulty to define properly scaled objective functions for each
criterion, such that a weighted sum really reflects the relative
importance the multiple criteria.

In order to avoid such issues we use a multi-objective optimization
approach~\cite{debMOOBook01,coelloMOOBook02,coelloTrendsMOO05,Ishibuchi_MAO_2008},
where a complete set of Pareto-optimal solutions is directly obtained.
The two objectives we focus on are the topological robustness of
networks against random and targeted removal of nodes.  These two types
of robustness are known to be in a trade-off relation, where increasing
the robustness with respect to one type of removal is likely to decrease
the
other~\cite{cohen_breakdown_2001,ValentePeak04,peixotoRobustTopology12}. In
particular, it has been recently shown that in absence of any
constraints other than a fixed average degree, the optimization of
robustness against random failure leads to a core-periphery structure,
where most nodes are connected to a core group, possessing a high
average degree, which is also internally
connected~\cite{peixotoRobustTopology12}. Although being maximally
robust against random failure, this core-periphery topology is minimally
robust against targeted attacks, since the removal of the few core
node  immediately leads to the vanishing of the giant component. This
robustness-fragility duality is a common feature of real networks with
heterogeneous structure; a famous example of which is the
Internet~\cite{doyle_robust_2005}.

In order to investigate this multi-objective optimization scenario, we
follow Ref.~\cite{peixotoRobustTopology12} and focus on large-scale
topological features, as parametrized by a stochastic block
model~\cite{holland_stochastic_1983, karrerBlockmodel11}. This
parametrization allows for arbitrary large-scale mixing patterns, such
as assortativity, dissortativity, community structure, core-peripheries,
etc., as well as arbitrary local degree distributions. This model is
also convenient, since it allows the exact computation of the
percolation properties of the system in the limit of large
networks~\cite{peixotoRobustTopology12, bujok_percolation_2014}.

By analyzing the Pareto-optimal fronts according to the two robustness
criteria, we observe that a minimal number of three blocks is sufficient
to obtain the optimal fronts, and that in most cases the best trade-off
is realized by a hybrid structure composed of a core-periphery and a
third ``secluded'' group, which is strongly connected internally and
marginally connected to the core nodes. The two-block core-periphery
of Ref.~\cite{peixotoRobustTopology12} and the fully random network are
recovered at the two extremes of the Pareto fronts, for maximum
robustness against random and targeted node removal, respectively.

This paper is organized as follows. In Sec.~\ref{sec:blockmodel}, we define
the stochastic block model and in Sec.~\ref{sec:robust} our
robustness criteria. In Sec.~\ref{sec:opt}, the evolutionary
multi-objective optimization algorithm is described briefly.
Sec.~\ref{sec:results} presents the results of the optimization for several
parameter choices, including the Pareto-optimal fronts
and the resulting topologies. In Sec.~\ref{sec:conclusion}, we finalize
with an overall discussion.


\section{Materials and Methods}
\subsection{Stochastic block model}
\label{sec:blockmodel}

The \emph{stochastic block model} defines an ensemble of random
networks, in which nodes belong to different groups (also called ``blocks''), and
the probability of an edge existing between nodes is a function of the
block membership of each node. Each block holds a fraction $n_r$ of the
$N$ nodes of the whole network, where $r\in[1, B]$ enumerates these
blocks and $B$ is the total number of blocks, such that
$\sum_{r=1}^{B}n_r=1$.  Following Ref.~\cite{karrerBlockmodel11},
each of the $B$ blocks is characterized by an independent degree
distribution $p_k^r$, which specifies the fraction of nodes with degree
$k$ in block $r$.

The connections between the blocks are described with a matrix
$\mathbf{e}$, where the elements $e_{rs}$ specify the number of
half-edges per node in block $r$ connecting to nodes in block $s$. For
simplicity of notation, the diagonal elements $e_{rr}$ encode twice the
number of edges per node within block $r$.

In the framework of the stochastic block model, the network structure
becomes locally tree-like when the number of nodes $Nn_r$ inside each
block is sufficiently large. Since the probability of an edge existing
between any two nodes of groups $r$ and $s$ scales as is
$Ee_{rs}/Nn_rNn_s \sim O(1/N)$, with $n_r \sim 1/B$ and $e_{rs} \sim
1/B^2$, the probability of an edge existing between any two chosen
neighbours will become vanishingly small as $N\to\infty$. Therefore,
since local substructures such as triangles are not generated by the
model, predictions based on block model calculations can only be
accurate for (large) tree-like networks without these local
substructures.  However, global and meso-scale properties such as
community structure~\cite{karrerBlockmodel11},
assortativity~\cite{newmanAssort03}, bipartite, core-periphery
structures~\cite{peixotoRobustTopology12}, or any other arbitrary mixing
pattern are well captured.

Each block of the network can in principle have an arbitrary degree
distribution $p_k^r$.  However, in this work we restrict the degree
distribution of each block to be a modified Poisson degree distribution,
\begin{equation}
  p^\m{MP}_k=(1-\delta_{k,0})\frac{\kappa^k}{(e^\kappa-1)k!},
  \label{eq:poisson}
\end{equation}
where $\delta_{ij}$ is the Kronecker delta function.
Thereby nodes with zero degree ($k=0$) are explicitly excluded, since they
can never belong to the giant component.  In
contrast to a regular Poisson distribution $p^\m{P}_k=\kappa^k
e^{-\kappa} / k!$, where the mean degree is directly given by $\kappa$,
the mean degree of the modified Poisson distribution is given by
$\km_\m{MP}=\kappa /(1-e^{-\kappa})$, which is always bigger than
$\kappa$. In particular, the mean degree cannot be less than one,
$\km_\m{MP}\geq 1$.

It is important to note that although the use of the modified Poisson distribution
as displayed in Eq.~\eqref{eq:poisson}
may seem like a strong imposition on the network structure, in reality
it is not. A large variety of nearly-arbitrary degree
distributions of the complete network can be obtained by composing many
blocks with different sizes and average degrees.

\begin{figure}[th!]
  \includegraphics[width=0.95\columnwidth]{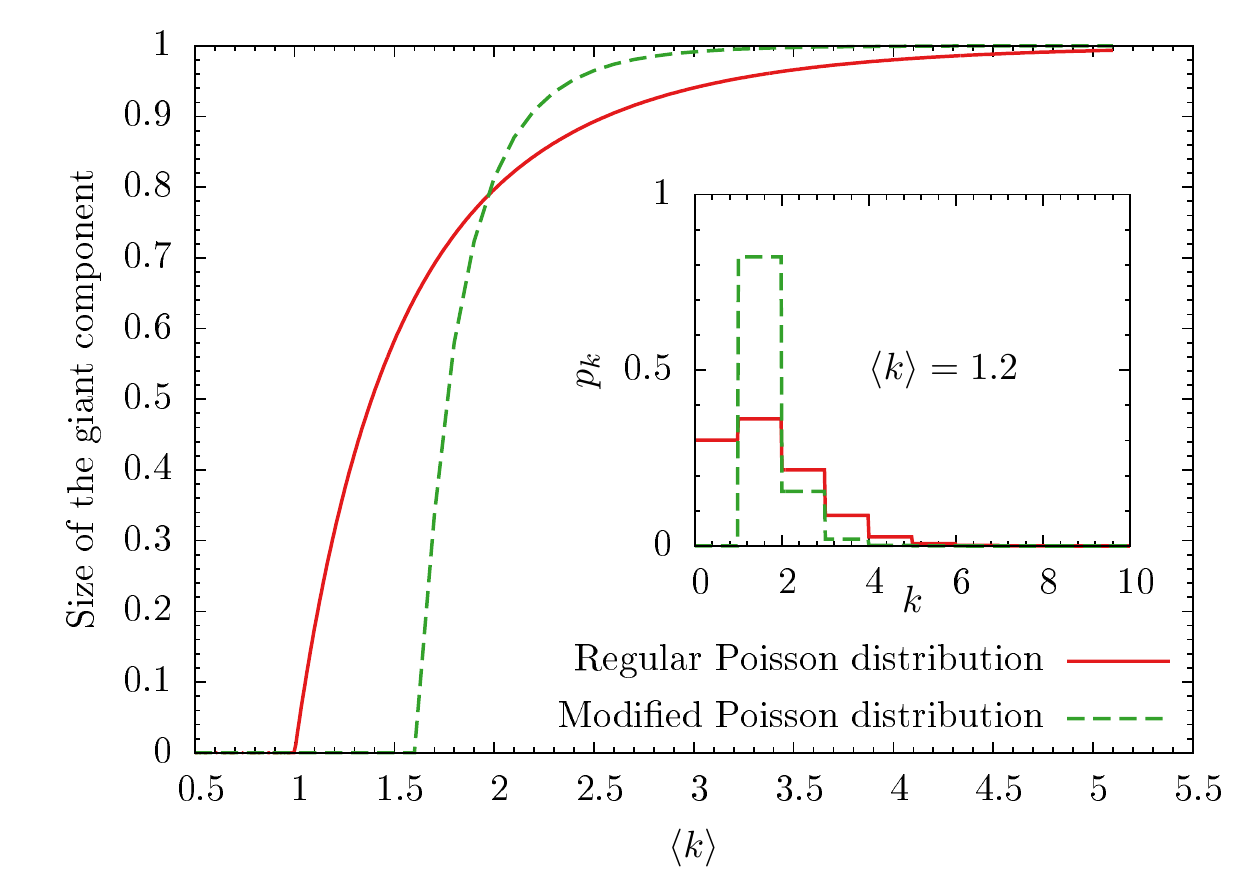} 
  \caption{{\bf Giant connected component of a Erd\H{o}s-R\'enyi random network with
    a Poisson (dashed green) and modified Poisson (solid red) degree
    distribution as function the average mean degree of the network.} The
    inset shows the regular and modified Poisson distribution for a mean
    degree of $\km=1.2$ with the same color coding.
    \label{fig:poisson}}  
\end{figure}

The percolation properties of a random network with the modified Poisson
degree distribution of Eq.~\eqref{eq:poisson} differ slightly from an Erd\H{o}s-R\'enyi
network, i.e.\ a random network with a regular Poisson distribution.  In
the Erd\H{o}s-R\'enyi network the percolation transition where a macroscopic connected
component emerges as a function of the mean degree occurs at
$\km_\m{P}^\m{C}=1$. In the case of the modified Poisson distribution,
the transition is shifted to $\km_\m{MP}^\m{C}=e/(e-1)\approx 1.58$, as
can be seen in Fig.~\ref{fig:poisson}.  This is a direct consequence of
the fact that no nodes with degree $k=0$ are allowed in the later case.
For a modified Poisson network to have low mean degree,
$\km_\m{MP}\gtrapprox 1$, a large fraction of nodes needs to have degree
one.  In order to achieve this, many of the $k-1$ nodes form pairs and
are thus isolated from the rest of the network.  Additionally, the
number of nodes with degree greater than one is strongly reduced compared
to the regular Poisson distribution (see inset of
Fig.~\ref{fig:poisson}).  This prohibits the existence of a macroscopic
connected component if the mean degree is close to one ($\km_\m{MP}\gtrapprox 1$).
Only when the fraction
of nodes with $k=1$ diminishes, a macroscopic giant component can form.
In this case, the nonexistence of disconnected $k=0$ nodes results in
larger connected components in general and leads to the stronger
increase of the size of the giant components as can be seen in
Fig.~\ref{fig:poisson}.

Apart from the degree distribution of each block, $p_k^r$, more
parameters need to be specified in order to define a realization of a
block model ensemble. These are the total number of blocks $B$, the
relative size of each block $n_r$, the mean degree of each block
$\km_r$, as well as the edges connecting the blocks given by $e_{rs}$.
These parameters are, however, not completely independent as the
relative sizes $n_r$ of all blocks must add up to one, $\sum_r n_r =1$,
and the sum of all the edges incident to one block is related to its
mean degree, $\km_r = \sum_s e_{rs}/n_r$.  Since we will always consider
networks with a given total mean degree \km, the following constraint
will need to be fulfilled, $\km=\sum_r n_r \km_r$.

\subsection{Node removal and robustness}
\label{sec:robust}

Failure in networks is modeled by removing a finite fraction $q$ of nodes
from the network. We will consider two different strategies for selecting which
nodes are removed. The first is \textit{random removal} where the nodes to be removed are
selected purely randomly. The second is \textit{targeted removal} 
where nodes with higher degree are more likely to be removed.

Both types of failures are inspired by real-world technical networks.
Random removal is considered to model fatigue of parts or other random
influences. Targeted removal is inspired by the fact that highly loaded
nodes are more likely to fail or, in the context of critical
infrastructure, malicious damage is preferably brought to important
nodes.

In the context of block models, where we only model representative nodes
in an statistical ensemble, we employ a slight variation of the targeted
removal which was also used in Ref.~\cite{peixotoRobustTopology12}.  The
targeted criterion is only applied to the selection of blocks where the
fraction of nodes to be removed from block $r$ is proportional to
$e^{\km_r}$ and thus increases with the mean degree of the block.
However, within each block no further targeted removal of nodes is
performed and nodes are removed at random.  In case of all blocks having
the same mean degree targeted removal is identical to random node
removal.

As a measure of robustness of a network we use the size of the
macroscopic component $S(q)$ after a finite fraction $q$ of nodes has
been removed.  Instead of focusing on the robustness when removing a
single fraction $q$, all possible values $0\leq q\leq1$ are considered
to obtain a sensible measure for the overall robustness of a network.
Therefore,  we define the robustness as it was proposed in Ref.~\cite{schneiderAttack11}
as
\begin{equation}
 R = 2\int_{0}^1 S\left(q\right)\;\text{d}q\quad\text{,}
\end{equation}
where the factor of 2 serves to adjust the range of $R$ to be
$\left[0,1\right]$.  The limiting case $R = 0$ is achieved by networks
without a macroscopic component, even when no nodes are removed at all.
The opposite limiting case of $R = 1$ requires a fully connected network
where $S(q) = 1-q$.

Following Ref.~\cite{peixotoRobustTopology12} using the generating function
formalism~\cite{newmanNetwork01} the size of the macroscopic component
is calculated using $u_r$, which is the probability that a node in block
$r$ is not connected to the macroscopic component via one of its neighbors.
These probabilities for all blocks have to fulfill a system of $B$
self-consistent coupled equations:
\begin{equation}
  u_r = \displaystyle\sum_{\substack{s}} m_{rs}
  \left[1+\frac{\phi_s}{\kappa_s}
    \left(
      g_{0,s}^{\prime}\left(u_s\right)-1
    \right)
  \right], \label{selfcon2}
\end{equation}
where $m_{rs} \equiv e_{rs}/n_r \kappa_r$ is the fraction of edges in
block $r$ leading to block $s$, $n_r$ and $\km_r$ are the relative
number of nodes and mean degree of block $r$, respectively, and
$g_{0,r}(z) = \sum_k p_k^r z^k$ is the generating function of the degree
distribution of block $r$ and
$g^\prime_{0,r}(z)=\frac{\partial}{\partial z} g_{0,r}(z)$ is its
derivative.  $\phi_r\in[0,1]$ is the fraction of nodes \emph{not}
removed from block $r$. The $\phi_r$ have to be chosen in accordance
with the node removal strategy, for example, $\phi_r=\phi $ for random
removal or $\phi_r \propto e^{-\km_r}$ for targeted removal. Since the
total fraction of removed nodes is given by $q$, the $\phi_r$ need to
satisfy the relation $q=1-\sum_s\phi_s n_s$. Due to this requirement,
the $\phi_r$ for targeted removal need to be determined by numerically
solving $0 = 1-q - \sum_r n_r\exp(-\km_r(1-x)/x)$ for $x$ and using
the solution $x^*$ to get $\phi_r = \exp(-\km_r(1-x^*)/x^*)$.

The solutions of these equations for all $u_r$ allows for the
calculation of the size of the giant connected  component  $S(q)$,
\begin{equation}
 S(q) = \displaystyle\sum_{\substack{s}} n_s\phi_s\left[1-g_{0,s}\left(u_s\right)\right].
\end{equation}

At this point, a few remarks about the interpretation of the value of
$S(q)$ should be made. Since we are parametrizing the system with
intensive
quantities ($e_{rs}$, $n_r$, $u_r$, etc.) which specify
\emph{fractions} of nodes and edges in infinitely large systems, we
cannot differentiate between the existence of single or multiple
macroscopic components for  a given value of $S(q)$.
In other words, if two macroscopic components
are connected by a single edge (or more generally, any intensive number of edges)
the probability of edges between them vanishes in the infinite size
limit. Thus, this situation cannot be distinguished from two truly
disconnected macroscopic components where no edges exists between the two
components.
For the purposes of this
work, we consider this issue to be unimportant, and we focus on the
existence of macroscopic components in the more abstract sense as given
by the value of $S(q)$ directly.

For each node removal strategy, Eqs.~\eqref{selfcon2} have to be solved for
all $q$ in order to calculate the robustness $R$ of a specific block
model ensemble. In our case, this leads to two different measures of
robustness, \ensuremath{R_\mathrm{Random}} and \ensuremath{R_\mathrm{Targeted}}, for random and targeted node removal,
respectively.

\subsection{Evolutionary optimization}
\label{sec:opt}

In order to consider both robustness measures, \ensuremath{R_\mathrm{Targeted}} and \ensuremath{R_\mathrm{Random}}, at once, we
utilize a
multi-objective~\cite{debMOOBook01,coelloMOOBook02,coelloTrendsMOO05,Ishibuchi_MAO_2008}
evolutionary optimization~\cite{baeckEvoComp97,simonEvoOpt13} algorithm.
Unlike in the optimization of a single objective, where it is always possible to
state if a certain solution $A$ is better, worse or equally good compared to a solution $B$, this is
not necessarily possible in multi-objective optimization.
If a solution $A$ performs better than a different solution $B$ in one
objective, but worse in a second objective, no statement is
possible which of the solutions is better.
Only if  solution $A$ is better than $B$ in at least one objective and not worse in any objective
it can be considered generally better and
it is then said that  $A$ \emph{dominates} $B$.
Sets in which no solution dominates any other solution are called
\emph{non-dominating}.
In general, a multi-objective
optimization will not result in a single best solution but in a set of
non-dominating solutions which ideally is close to the best possible set
of non-dominating solutions, the Pareto-optimal front.
These non-dominating sets are very useful to
study the trade-off relation between the robustness \ensuremath{R_\mathrm{Targeted}} and \ensuremath{R_\mathrm{Random}} and
their relation to the structure of optimal networks.

The algorithm we use here is the so called \textit{S-metric selection
evolutionary multi-objective optimization algorithm}
(SMS-EMOA)\cite{beumeSMSEMOA07}. It is a population based evolutionary
stochastic search algorithm which does not utilize any gradient
information and is well suited for non-convex and noisy optimization
problems.  The algorithm does not optimize the objectives directly but
instead maximizes the hypervolume in objective-space dominated by the
population and bound by a reference point.  In the present case of two
objectives, the hypervolume is given by the area under the Pareto-curves
as, for example, shown in Fig.~\ref{fig_pareto_2blocks}.  At each
iteration the solution whose removal leads to the lowest decrease in the
dominated hypervolume is removed from the population and a new solution
is generated by recombination and mutation (for more details
see~\cite{deb1994simulated}).

Repeating the steps of removing the least contributing solution and
generating a new solution not only shifts the solution set closer to
the Pareto-optimal front but also leads to a broad distribution along
the front, two desired properties of an optimal set of
solutions.\footnote{
  For completeness, we state the parameters used for the SMS-EMOA: A
  population size of 50 is used, the
  crossover probability is $p_\mathrm{c}=1$, the crossover distribution parameter is
  $\eta_{m}=20$,
  the mutation probability is $p_\mathrm{m}=1$, and the mutation distribution parameter is $\eta_{c}=15$.
}

For each optimization run, we fix the number of blocks $B$ and the mean
degree of the complete network, $\km$.  Each block has a modified Poisson
distribution as its degree distribution (cf.\ Eq.~\eqref{eq:poisson}),
but the average mean degree of each block can vary.  Therefore, the free
variables subject to optimization, i.e.\ the search parameters, are the
relative size of each block, $n_r$, the mean degree of each block,
$\kappa_r$, and the entries in the matrix containing edges within and
between the blocks, $e_{rs}$.  With the sum rules and constraints stated
at the end of Section \ref{sec:blockmodel}, this results in $\frac12
B(B+1)+B-2$ independent search parameters.

\section{Results}
\label{sec:results}

\subsection{Trade-off curves}
\label{sec:pareto}

In Fig.~\ref{fig_pareto_2blocks} we show
robustness values obtained from
different optimizations for several numbers of blocks
($B={2,3,4,5}$), but all with a fixed  mean degree of $\km=2.5$.
The Pareto-optimal fronts for optimizations with $B=3$, $4$, and
5 blocks match exactly.  Only the $B=2$ result deviates
and yields lower robustness over large parts of the
Pareto-optimal front.

\begin{figure}[th!]
   \centering
   \includegraphics[angle=0, width=0.9\columnwidth]{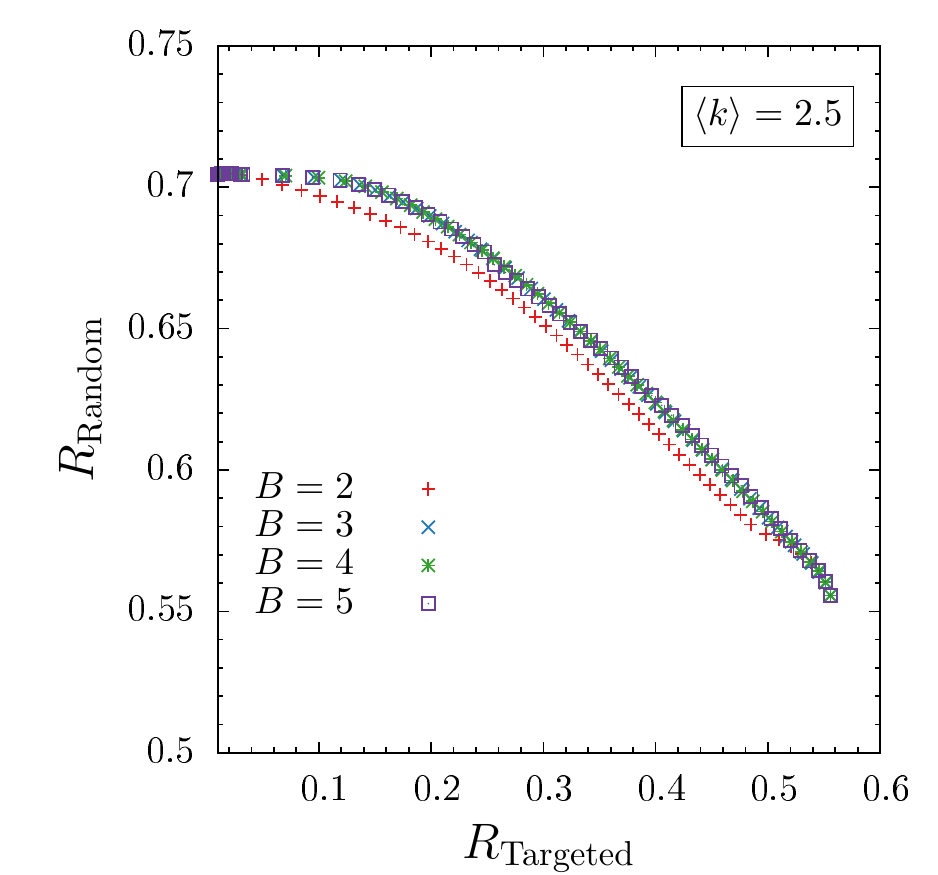} 
   \caption{{\bf Pareto-optimal fonts of robustness against targeted and random
     removal of nodes for mean degree $\km = 2.5$ and
     various  number of blocks.} 
      \label{fig_pareto_2blocks}}
\end{figure}

The network structures corresponding to the Pareto-optimal solutions for
$B=3$, $4$, and $5$ blocks are also identical (not shown)~\footnote{Two
structures with different $B$ values are considered identical when their
structural entropy is the same. See Sec.~\ref{sec:structure} for more
details.}.  The same behavior was found for other values of the mean
degree $\km$, where the results for $B\geq 3$ were identical and
deviations were only observed for $B=2$.

This leads to the conclusion that three blocks are sufficient to
describe networks which are maximally robust against random and targeted
node removal.  At both extremes of the Pareto-optimal fronts all curves
coincide, which means that for optimizing only with respect to one
objective (i.e. single-objective optimization),
$B\le 2$ is sufficient to achieve maximal robustness (see Section
\ref{sec:structure}).  This is in accordance with the results of
Ref.~\cite{peixotoRobustTopology12}, where single-objective optimization
was performed, and a $B=2$ core-periphery and a $B=1$ fully random
structures were found as optimum for random and targeted node removal,
respectively. This is also consistent with the findings of Valente et
al.~\cite{ValentePeak04} who showed two- and three-peak degree
distributions to be optimal when minimizing percolation thresholds of
networks  subject to random and targeted removal of nodes. 

\begin{figure}[th!]
  \includegraphics[width=0.95\columnwidth]{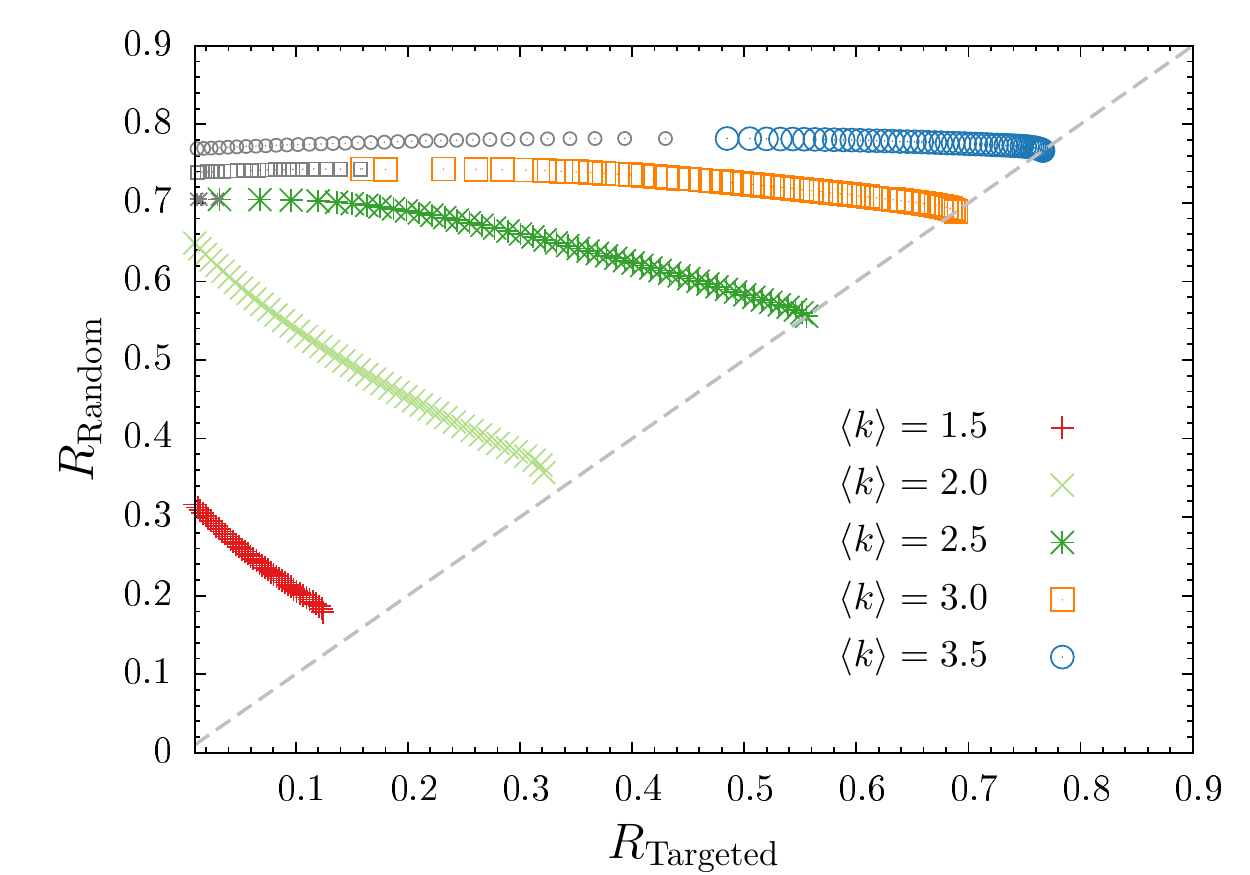} 
  \caption{{\bf Pareto-optimal fronts of \ensuremath{R_\mathrm{Random}} versus \ensuremath{R_\mathrm{Targeted}}
    for optimal block model networks with $B=3$  and for various mean
    degrees (colored symbols).} 
    For $\km > 2$, the smaller gray symbols to the left of
    each Pareto front
    indicate solutions  which
    maximize \ensuremath{R_\mathrm{Random}} for fixed \ensuremath{R_\mathrm{Targeted}} but which are
    not Pareto-optimal (see main text).
    \label{fig_allFronts}}  
\end{figure}
The Pareto-optimal fronts of optimized block model networks with three
blocks ($B=3$) and a mean degree $\km$ between $1.5$ and $3.5$ are shown
in Fig.~\ref{fig_allFronts}. As intuitively expected, the general
trend where the robustness increases with the mean degree is observed.

For small values of the mean degree, $\km\lesssim 2.5$, the two types of
robustness are in strong trade-off relation: Increasing the robustness
against targeted removal strongly decreases the robustness against
random removal (and vice-versa).

Pareto-optimal solutions of networks with $\km\lesssim 2$ and with the
highest robustness against random removal are always found to be most
fragile with respect to targeted removal, i.e.\ at $\ensuremath{R_\mathrm{Targeted}}=0$.  But for
$\km>2$, networks with the maximal value of \ensuremath{R_\mathrm{Random}} shift to have a finite
robustness against targeted removal, $\ensuremath{R_\mathrm{Targeted}}>0$.  In this case, networks
with lower robustness against targeted removal are not accessible via
the multi-objective optimization, since they are not Pareto-optimal
(i.e. they are dominated by the solutions with maximal \ensuremath{R_\mathrm{Random}}, see
Refs.~\cite{debMOOBook01,coelloMOOBook02,coelloTrendsMOO05,Ishibuchi_MAO_2008}).
However, they can be found by performing an optimization with the
value of \ensuremath{R_\mathrm{Targeted}} fixed, and such results are shown as the smaller gray symbols in
Fig.~\ref{fig_allFronts}.  The Pareto optimal front together with these
additional solutions form the whole trade-off curve for each $\km$.

With increasing mean degree, the trade-off curves become very flat,
indicating that a slight sacrifice on the robustness with respect to
random removal yields a great enhancement in the robustness against
targeted removal.  Additionally, the curves increasingly approach the
diagonal where $\ensuremath{R_\mathrm{Random}}=\ensuremath{R_\mathrm{Targeted}}$, which means that there are solutions which
are equally good in
both measures.

In general \ensuremath{R_\mathrm{Random}} is always greater or equal to \ensuremath{R_\mathrm{Targeted}}, and for $\km \gtrsim
2.5$, the Pareto-optimal fronts extend to the diagonal.
In random networks, nodes with high degree are important for the size of
the giant component since they naturally are more likely to connect
different components.  Due to this, a removal mechanism targeting high
degree nodes is able to degrade the giant component easily by removing a
relatively small amount of high degree nodes.  Therefore, making the
degree distribution of a network narrow should increase the robustness
against targeted removal since there are less high-degree nodes.  In a
block model with several blocks a narrow degree distribution implies
that all blocks have the same mean degree $\km_r = \km$.  Since, in this
work, targeted removal only differentiates between blocks but not
between nodes inside the block, targeted and random removal are
identical if all blocks have the same mean degree.  As a consequence,
the robustness values are then equal, $\ensuremath{R_\mathrm{Random}} = \ensuremath{R_\mathrm{Targeted}}$.

In contrast, for $\km \lesssim 2.0$, the Pareto-optimal fronts do not
extend to the diagonal, which is a consequence of the percolation
properties of fully random $B=1$ networks (cf.\ Section
\ref{sec:blockmodel}).  For low mean degrees, the giant connected
component of a fully random network is very small even without node
removal ($q=0$).  Due to the steep increase of the giant component with
increasing mean degree (cf.\ Fig.~\ref{fig:poisson}), it is beneficial
to have two blocks with differing mean degree, one higher and the other
lower than the total average mean degree \km.  The block having a mean
degree greater than $\km$, also has a substantial larger giant component,
while the giant component of the other block is still small (or even
zero).  Therefore, the argument presented above for $\km \gtrsim 2.5$,
where a finite giant component at $q=0$ always exists, is not effective
for $\km\lesssim 2$.  It is always beneficial to have (at least) two
blocks in order to have increase the size of the giant component for
$q=0$.

\subsection{Network structures}
\label{sec:structure}

In our approach, the number of blocks $B$ is set a priori and kept fixed
during a single optimization procedure. However, networks
with different values of $B$ could have equivalent topologies. This can
happen if one or more blocks have a vanishing size $n_r$ and mean degree
$\km_r$, or when two or more blocks can be merged together without
altering the ensemble of generated networks.

For a clearer visualization and analysis of block model structures, we
reduce the number of blocks by removing insignificant blocks and by
merging multiple blocks into one if they are equivalent.  For two blocks
to be equivalent, we require that the entropy of the merged and the
original network ensembles differ by a very small amount. The entropy of
the stochastic block model ensemble is simply the logarithm of the total
number of networks which can be generated given a specific
parametrization, i.e. choices of $n_r$ and $e_{rs}$. The entropy is a
signature of the ensemble, and determines how random it is. If the
entropy remains the same after two blocks are merged into one, this
means that these two groups correspond simply to an arbitrary
subdivision of a larger group, and they do not in fact constrain the
topology in any way. The entropy of block model networks is calculated
as described in Ref.~\cite{peixotoBlockmodels12}. We emphasize that,
since the topologies in this case are in fact equivalent, the effect of
the merging process on the robustness values was found to be negligible.

We now consider the Pareto fronts separately for different values of
the mean degree.

\subsubsection{Networks with intermediate mean degree $\km = 2.5$}

In Figure \ref{fig_k2.5_3Blocks}, the structure and parameters of
optimized networks for $\km = 2.5$ are depicted.  The merging procedure
is reflected in the fact that the number of blocks indicated by the
number indices on the axis and the number of possible squares in the
top-row Hinton-plots varies between one and three.

\begin{figure}[th!]
  \centering
 \includegraphics[width=\columnwidth]{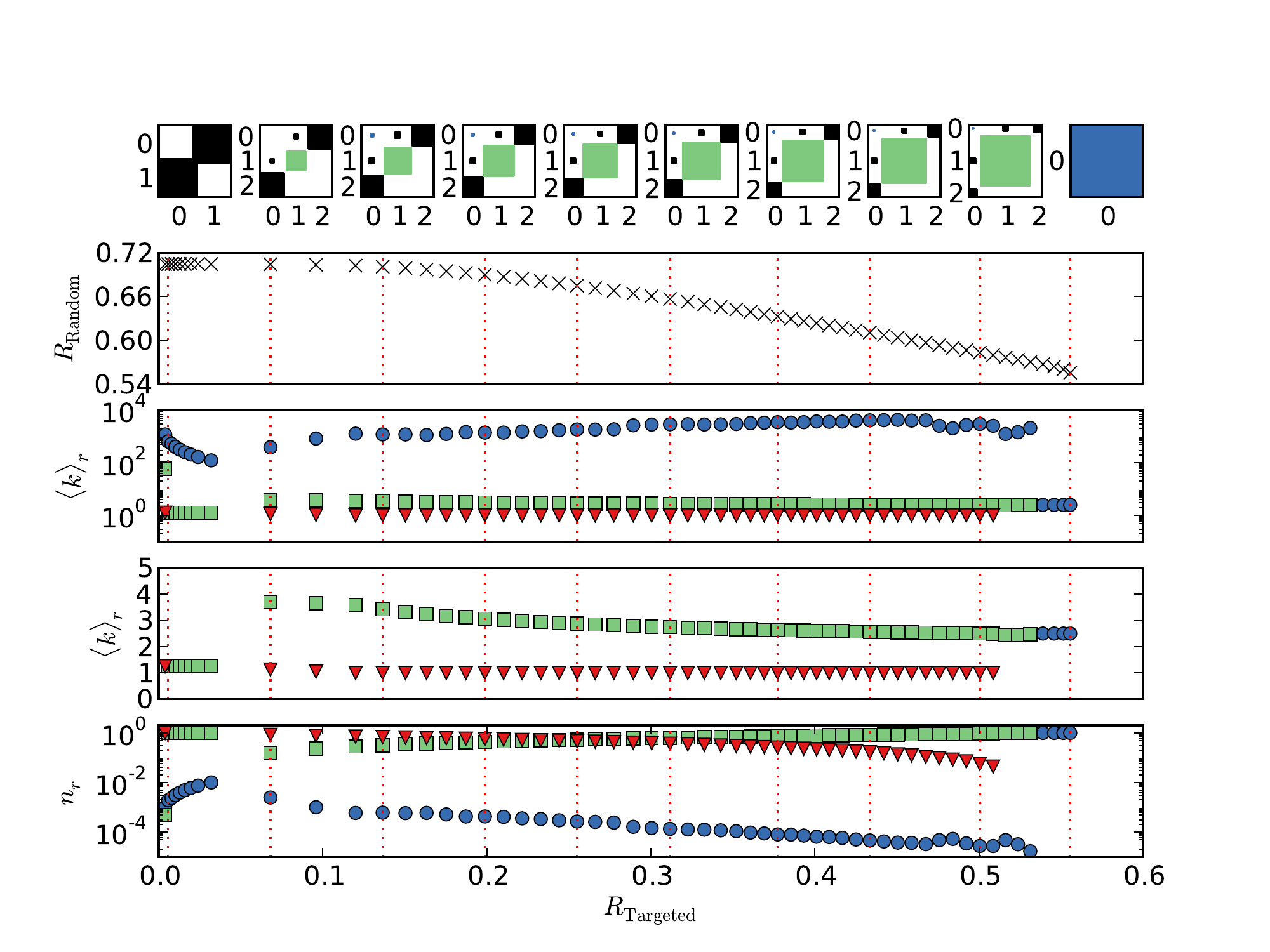} 
  \caption{{\bf Parameters of the optimized networks as a function of \ensuremath{R_\mathrm{Targeted}}
    obtained from a three-block optimization with $\km = 2.5$.}
    The upper row shows the elements of the edge matrix $e_{rs}$,
    where the areas of the squares is proportional to the logarithm of the element. 
    The positions for which these Hinton plots are shown are marked with
    dashed lines in the other panels. The second row shows the trade-off curve
    already displayed in Fig.~\ref{fig_pareto_2blocks},
    while the third and fourth display the mean degree of the blocks, 
    on a logarithmic and a linear scale, respectively.
    The last row shows the relative sizes of the blocks.
    The coloring  of the blocks and their index is determined by their mean degree,
    where the block with the highest mean degree is shown in blue and always has
    index $r=0$, followed by green with $r=1$ (second highest) and red
    ($r=2$,  lowest degree).
    \label{fig_k2.5_3Blocks}
}  
\end{figure}

For $\ensuremath{R_\mathrm{Targeted}}\to 0$ we recover the core-periphery structure found in
Ref.~\cite{peixotoRobustTopology12} where the optimal solution consists
of only two blocks.  One block is the very large periphery block which
contains nearly all the nodes ($n_\per\approx 1$) and which has the
lowest mean degree possible in this kind of structure $\km_\per \approx
\km/2= 1.25$.  The core block contains only very few ($ n_\co\sim
10^{-3} $) but very high-degree nodes ($\km_\co\sim 10^3$).  Almost all of the
edges are between the core and the periphery.

The core is central for forming the giant component, but takes up only a
very small fraction of the network.  Therefore, random removal will
almost always affect periphery nodes and the giant component will shrink
approximately linearly with the number of removed nodes, which is as
slow as possible.

On the other hand, the core-periphery structure is 
maximally fragile with respect to targeted removal, since
removing the core  completely removes the giant component.

With increasing robustness against targeted
removal, a third block emerges for $\ensuremath{R_\mathrm{Targeted}}\gtrsim 0.07$
in addition to the core and periphery block.
This new block, which we will call the \textit{secluded block}
is first of medium size ($n_\secl\approx 0.16 $)
and has a mean degree of $\km_\secl\approx 4$.
In contrast to the core and the periphery block,
it has a substantial amount of edges internally,
i.e.\ edges between nodes within this block (green square in the Hinton plots).
The secluded block is only lightly connected to the core block
and no edges exist between secluded  and periphery block.

Increasing \ensuremath{R_\mathrm{Targeted}} further, the
mean degree of the secluded block slightly decreases,
while it grows in size. The number of nodes in the
periphery continuously decreases and around
$\ensuremath{R_\mathrm{Targeted}}\approx 0.24$ the secluded block is larger
than the periphery.
For very high $\ensuremath{R_\mathrm{Targeted}}\gtrsim 0.52$ the secluded block dominates
and the core periphery structure vanishes.
The result is a single block network with a modified Poisson degree distribution,
as it was already mentioned  in the discussion at the end of  Section\ \ref{sec:pareto}
in connection
with Fig.~\ref{fig_allFronts}. 

Considering the complete Pareto-optimal set of solutions,
the dominant structure is a
three-block structure with a small but very high
degree core, a large but low degree periphery and an additional secluded block which
has a medium mean degree. Connections
only exist between the core and the periphery, and between the core and
the secluded block. The structure is best qualified as a modified core-periphery
with a regular Erd\H{o}s-R\'enyi network attached to the core. The relative size of the
secluded Erd\H{o}s-R\'enyi  block compared to the core-periphery structure  grows with an
increased robustness against targeted node removal.

\subsubsection{Networks with low mean degree $\km = 2$}

\begin{figure}[th!]
   \centering
   \includegraphics[angle=0, width=\columnwidth]{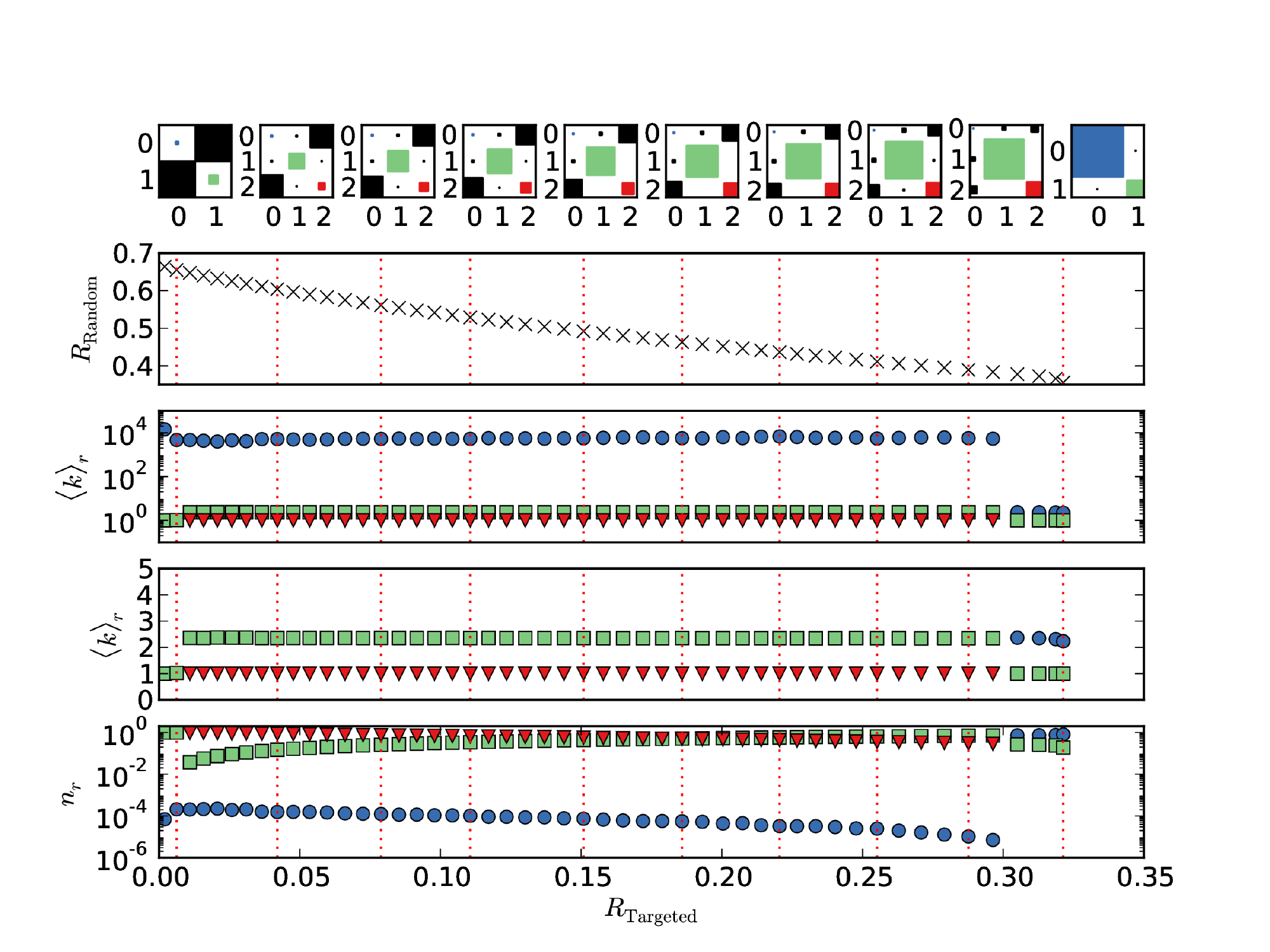} 
   \caption{{\bf Parameters of the networks along the
     trade-off curve for the five block optimization with $\km =
     2.0$.}  The panels are the same as in Fig.\ref{fig_k2.5_3Blocks}
     \label{fig_k2.0_3Blocks}
   }
\end{figure}
The structures of Pareto-optimal networks with a low mean degree of $\km = 2$
are shown in Figure \ref{fig_k2.0_3Blocks}.
The resulting structures are overall quite similar
to the previously discussed case with $\km=2.5$.
For $\ensuremath{R_\mathrm{Targeted}}\to0$ a core-periphery structure results, with
an additional secluded Erd\H{o}s-R\'enyi block emerging as \ensuremath{R_\mathrm{Targeted}} is increased. 

However, a striking difference to the situation for $\km = 2.5$
is that the number of edges within the periphery does not
vanish but is finite for all structures. 

The periphery block always has a mean degree very close to one, $\km_\per\gtrsim1$,
which implies that the majority of nodes has
exactly degree one. Therefore, most edges within the periphery
produce an isolated pair of two nodes not connected to
any macroscopic component (cf.\ discussion of the modified Poisson
distribution in Section \ref{sec:blockmodel}).

At first glance, this seems contradictory as the giant component is
already reduced without any node removal ($q=0$).  However, this is
beneficial for the overall robustness as it allows for the rest of the
nodes to have a higher mean degree putting it further above the
percolation threshold.  As can be seen from Fig.~\ref{fig:poisson}, this
is especially effective for increasing the macroscopic component of the
secluded block as its mean degree of $\km_\secl\approx 2.3$ is close to
the steep increase in the size of the giant component.  On the other
hand, this may be viewed simply as an artifact of the specific
constraints we have imposed. Perhaps a more realistic scenario would be
to impose additionally that the size of the largest component cannot
decrease for any value of $q$ after the optimization. However, this
would make the analysis significantly more complicated, and would only
affect the outcome of very sparse networks.

Interestingly, this holds for the two-block core-periphery structure as
well for $\ensuremath{R_\mathrm{Targeted}}=0$.  A pure core-periphery
structure is especially expected for $\km=2$, since then the extreme
topology of a star can be realized (the Hinton of which plot is not
shown in Fig.~\ref{fig_k2.0_3Blocks}).  However, a very slight increase
in the robustness against targeted removal to
$\ensuremath{R_\mathrm{Targeted}}\approx 0.006$, leaves the two-block
core-periphery structure intact, but produces a significant amount of
pairs in the periphery (see leftmost Hinton plot of
Fig.~\ref{fig_k2.0_3Blocks}).  The size of the core jumps from
$n_\co\approx 7\times 10^{-5}$ to $n_\co\approx 20 \times 10^{-5}$ while
its mean degree is reduced from $\km_\co\approx 14 \times 10^3$ to
$\km_\co\approx 4.6 \times 10^3$.  With this structural change, a little
robustness against random removal is lost, but the a finite number of
edges is realized within the core which provides a finite robustness
against targeted removal.

As expected from the discussion at the end of Section~\ref{sec:pareto},
the structure with a maximal \ensuremath{R_\mathrm{Targeted}} consists of two blocks, one with a mean
degree $\km_0 > \km$ and another with $\km_1 < \km$.  Both blocks in
fact form largely independent components since there are very few connections
between them. This is a situation similar to the ``onion-like''
structure found in Ref.~\cite{schneiderAttack11} when optimizing against
targeted node removal while preserving a heterogeneous degree sequence.
There,
the nodes with higher degree are kept isolated from the rest of the
network, hence effectively functioning simply as ``bait'' for the
targeted removal, whereas the rest of the system remains intact.

It is very remarkable that for $\km=2$ the mean degree of the three
blocks stay constant over the complete range of the Pareto-optimal front
(apart from the far extremes). The optimal trade-off between the two
robustness measures can be achieved by only changing the connection matrix and the relative sizes of the blocks.

\subsubsection{Networks with high mean degree $\km = 3.5$}

\begin{figure}[th!]
   \centering
  \includegraphics[width=\columnwidth]{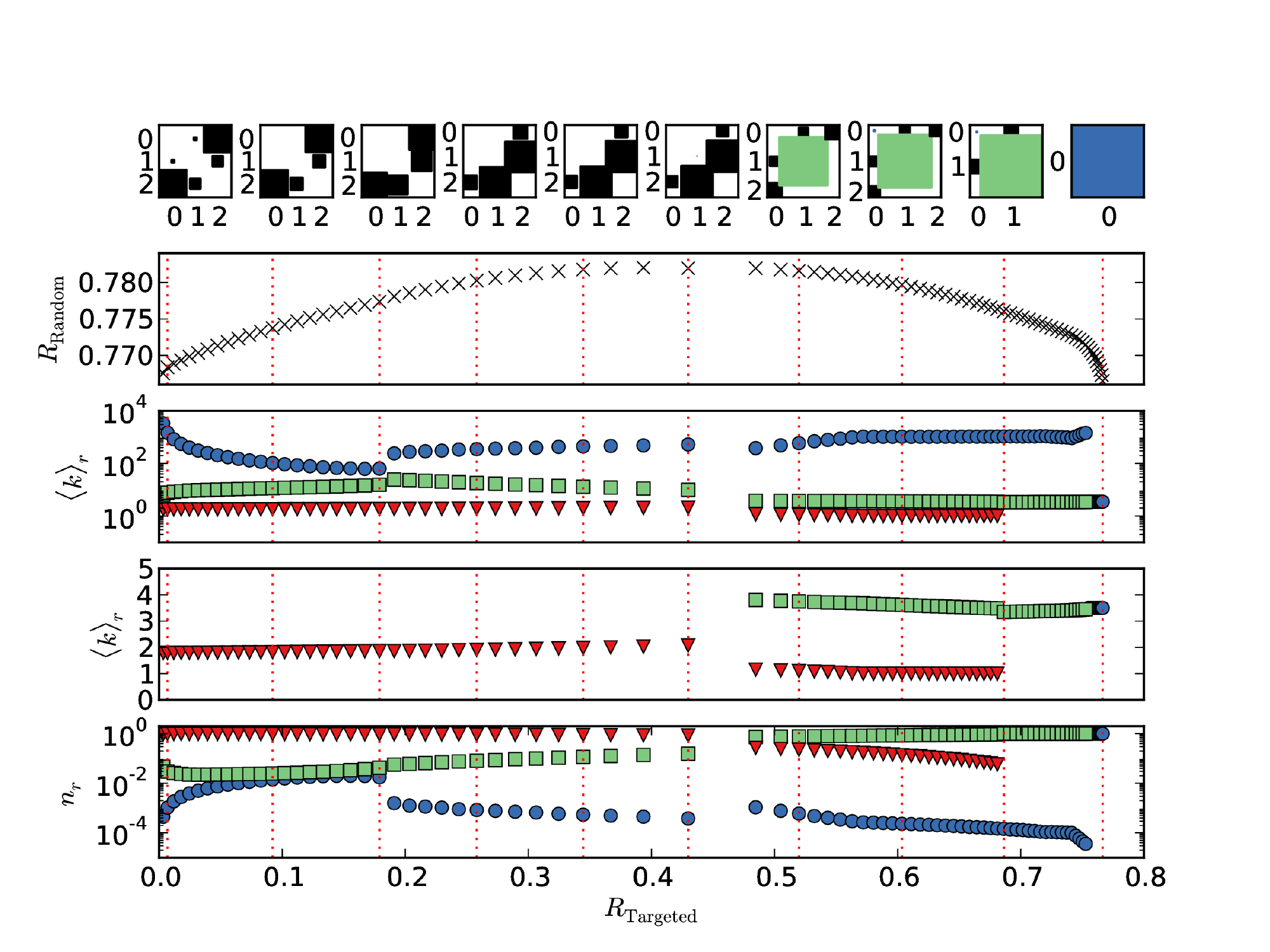} 
   \caption{{\bf Parameters of the networks along the
     trade-off curve for the five block optimization with $\km =
     3.5$.} The panels are the same as in Fig.\ref{fig_k2.5_3Blocks}.
     \label{fig_k3.5_3Blocks}}  
\end{figure}

The network structures for a higher mean degree of $\km=3.5$ is shown in
Fig.~\ref{fig_k3.5_3Blocks}. The Pareto-optimal part of the front,
that is for $\ensuremath{R_\mathrm{Targeted}} \gtrsim 0.48$, displays the same three-block structures
as for $\km=2.5$ and also reduces to a single block for maximal \ensuremath{R_\mathrm{Targeted}} where $\ensuremath{R_\mathrm{Targeted}}=\ensuremath{R_\mathrm{Random}}$
(cf.\ Fig.~\ref{fig_k2.5_3Blocks}).

For the part of the trade-off curve to the left of the maximum
of $\ensuremath{R_\mathrm{Random}}$  (i.e.\ for  $\ensuremath{R_\mathrm{Targeted}} \lesssim 0.42$), the structures  
change significantly. A three-block structure prevails
but the secluded block ceases to exist. No block has a
significant amount of internal edges and all edges connect
different blocks.
The largest block incorporates most of the nodes, $n\gtrsim 0.9$,
and has the lowest mean degree of $\km\approx2$.
A second block is very small with $10^{-5}<n<10^{-2}$
and has a high degree $10^{2}<\km <10^{4}$
and therefore strongly  resembles the core block.
The third block is of intermediate size
and degree, $10^{-2}< n <10^{-1}$ and $10<\km <10^{2}$, respectively.
The midsized and the small block are only connected via the
largest block since there are no direct edges between them.

For very low robustness against targeted removal, $\ensuremath{R_\mathrm{Targeted}}\gtrsim 0$, most
of the edges are between the core and the
largest block with $\km=2$.
With increasing robustness against targeted removal
the number of edges between core and the $\km=2$ block decreases
while more edges emerge between the $\km=2$ block and the midsized
block. At around $\ensuremath{R_\mathrm{Targeted}}\approx 0.19$ the same number of edges exist from
the $\km=2$ block to both of the other two blocks.
For higher \ensuremath{R_\mathrm{Targeted}}
  more edges exist between the $\km=2$ block and the midsized block.

This structural evolution can be understood by noting that the largest
part of the network always has a mean degree very close to two and acts
a connecting layer between the core and the midsized block.  For low
\ensuremath{R_\mathrm{Targeted}}, very few edges are between the block with $\km=2$ and the
midsized block, so that a connecting path between two different nodes
of the midsized block is very likely to traverse one of the few core
nodes.  Therefore, removing the core quickly fragments the network into
small components. On the other hand,  increasing the number of edges
between the $\km=2$ and the midsized block, a connecting path between
nodes within the midsized block is more likely to involve no nodes from
the core.  The core becomes increasingly unimportant and therefore the
robustness with respect to targeted removal increases.

\section{Conclusion}
\label{sec:conclusion}

In this paper we investigated the trade-offs between topological
robustness of networks against random and targeted node removals. We
used the stochastic block model to parametrize arbitrary mixing
patterns, and a multi-objective optimization algorithm to obtain the
Pareto-optimal fronts.  It was found that in order to achieve a
Pareto-optimal combination of robustness against random and targeted
removal, a network composed of at most three different blocks is
sufficient. In many cases the networks along the Pareto-optimal fronts
are composed of a hybrid topology, comprised of a core-periphery
structure, in addition to a secluded group, which is only sparsely 
connected to the core of the network, and not at all with the periphery.

At the edges of the Pareto-fronts, where one of the two robustness criteria
is maximized, one or two blocks suffice to obtain optimal networks:
A two-block structure is maximally robust against random failure,
and a fully random network with one block is sufficient in the case of targeted
removal. This reproduces
the results of Ref.~\cite{peixotoRobustTopology12}, and is also
consistent with the earlier findings of Valente et
al.~\cite{ValentePeak04} who found two- or three-peak degree
distributions to be optimal when minimizing percolation thresholds, with
networks which are otherwise fully random.

For low mean degrees of the overall network, the optimal robustness
values are generally lower than for higher mean degrees and a
significant trade-off exists between robustness against random and
targeted removal. With increasing mean degree the strong trade-off
diminishes and it is increasingly possible to obtain a high robustness
with respect to both criteria.  This implies that a network optimized
against one type of failure does not necessarily lose much of its
robustness when it is subsequently optimized against the other type of
failure or attack. Hence this shows that increasing the overall
connectivity of the network not only has the expected trivial effect of
increasing each robustness criterion individually, but to a large extent also allows for
them to be fulfilled simultaneously. This suggests
that the simple strategy of increasing the total number of edges in the
network, if combined with the optimal large-scale structures present in
the Pareto-optimal front, can be much more beneficial than could be
expected otherwise.

A comparison of the large-scale structures we find with the ones
observed in empirical
systems~\cite{albert_structural_2004,doyle_robust_2005,verma_revealing_2014}
is a natural and important extension of this work, and one we intend to
pursue in the future. The most appropriate approach is to search for
precisely the same type of model we are using in the analysis, which can
be done by inferring the parameters of the stochastic block model itself
from empirical data, which is a very active field of
research~\cite{holland_stochastic_1983,
karrerBlockmodel11,decelle_inference_2011,peixoto_parsimonious_2013}. In
fact, core-periphery structures have already been detected, such as the
topology of the internet at the autonomous systems level presented
recently in~\cite{peixoto_hierarchical_2014}.  However, to our
knowledge, an empirical verification of the specific structures we have
found has not yet been made.

In this work we have considered maximally robust networks that are
obtained when very few constraints are imposed. This gives us
fundamental limits on the multi-objective optimization against random
failures and targeted attacks. However, in empirical systems, exogenous
constraints are almost always present, such as geographical confinement,
and restrictions due to functional performance. In previous
studies~\cite{HerrmannOnion11, schneiderAttack11}, the optimization
against targeted node removal was considered when the degree sequence of an
empirical network is preserved. It was found that an assortative
structure emerges as the optimum in this case, where nodes become
connected with other nodes with similar degree. This has been obtained
as well by imposing similar constraints with the block model approach in
Ref~\cite{peixotoRobustTopology12}. However, it is still unknown how the
multi-objective optimization would behave when these constraints (and
other more realistically motivated ones) are simultaneously imposed. We
leave these questions for future work.

\newpage
\section*{Supporting Information}

\begin{figure}[th]
   \centering
  \includegraphics[angle=0, width=0.9\columnwidth]{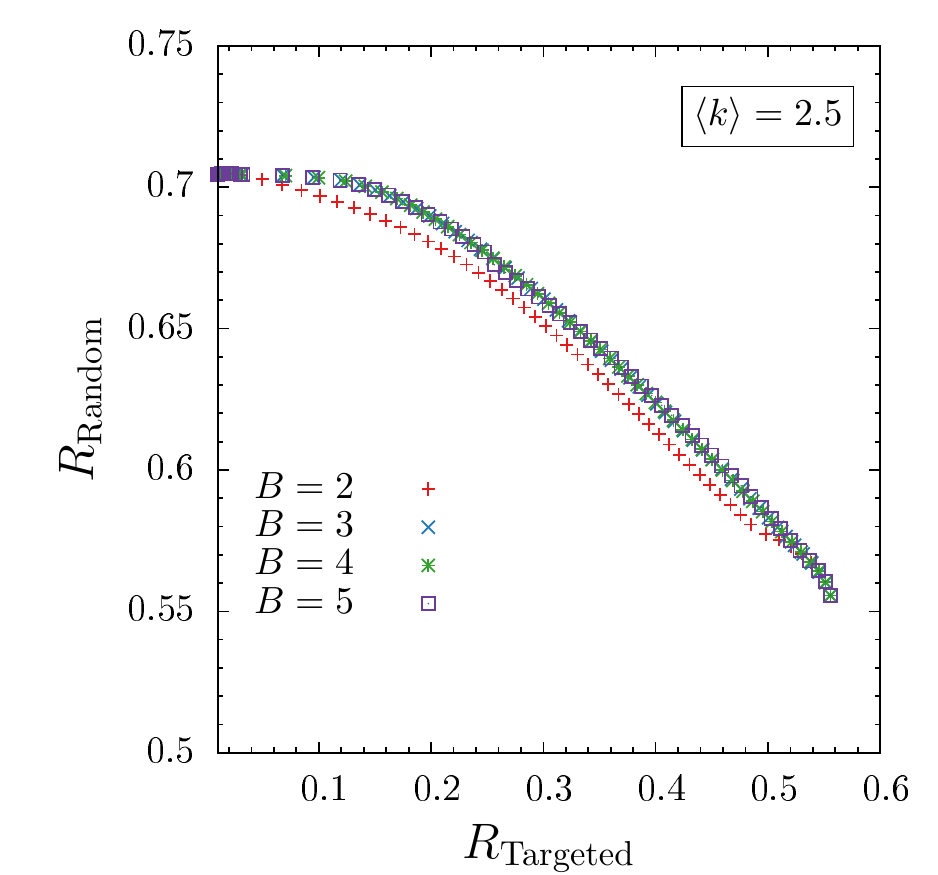} 
   \caption{{\bf Pareto-optimal fonts of robustness against targeted and random
     removal of nodes for mean degree $\km = 2.5$ and
     various  number of blocks.} 
     For three, four and five blocks the curves match
     exactly which implies that no more than three
     blocks are necessary to achieve the best robustness
     values. Since at the and of the curves all of
     them match only two or even one block is enough
     to achieve best robustness.
      \label{fig_pareto_2blocks}}
\end{figure}

\begin{figure}[th]
 \includegraphics[width=0.95\columnwidth]{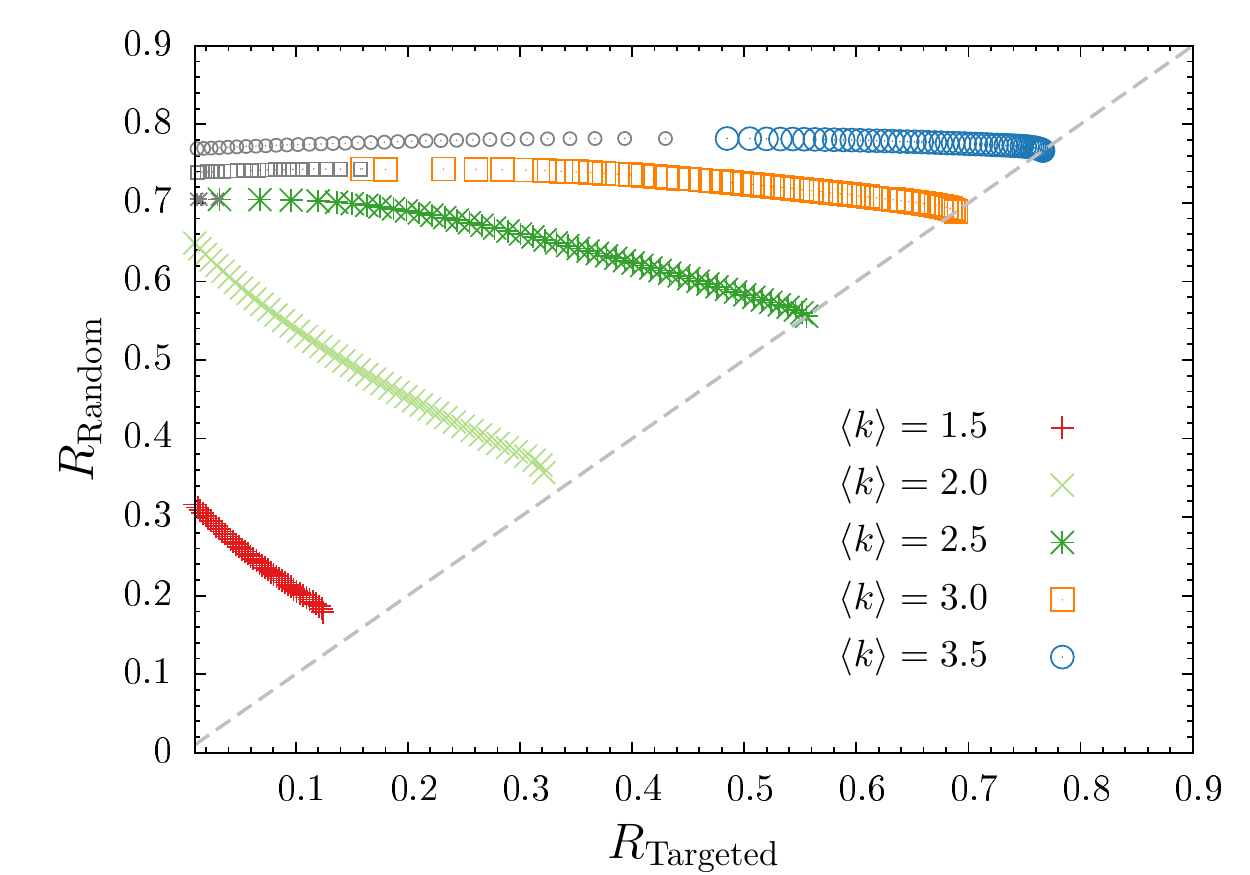} 
  \caption{{\bf Pareto-optimal fronts of \ensuremath{R_\mathrm{Random}} versus \ensuremath{R_\mathrm{Targeted}}
    for optimal block model networks with $B=3$  and for various mean
    degrees (colored symbols).} 
    For $\km > 2$, the smaller gray symbols to the left of
    each Pareto front
    indicate solutions  which
    maximize \ensuremath{R_\mathrm{Random}} for fixed \ensuremath{R_\mathrm{Targeted}} but which are
    not Pareto-optimal (see main text).
    \label{fig_allFronts}}  
\end{figure}

\begin{figure}[th!]
  \centering
  \includegraphics[width=\columnwidth]{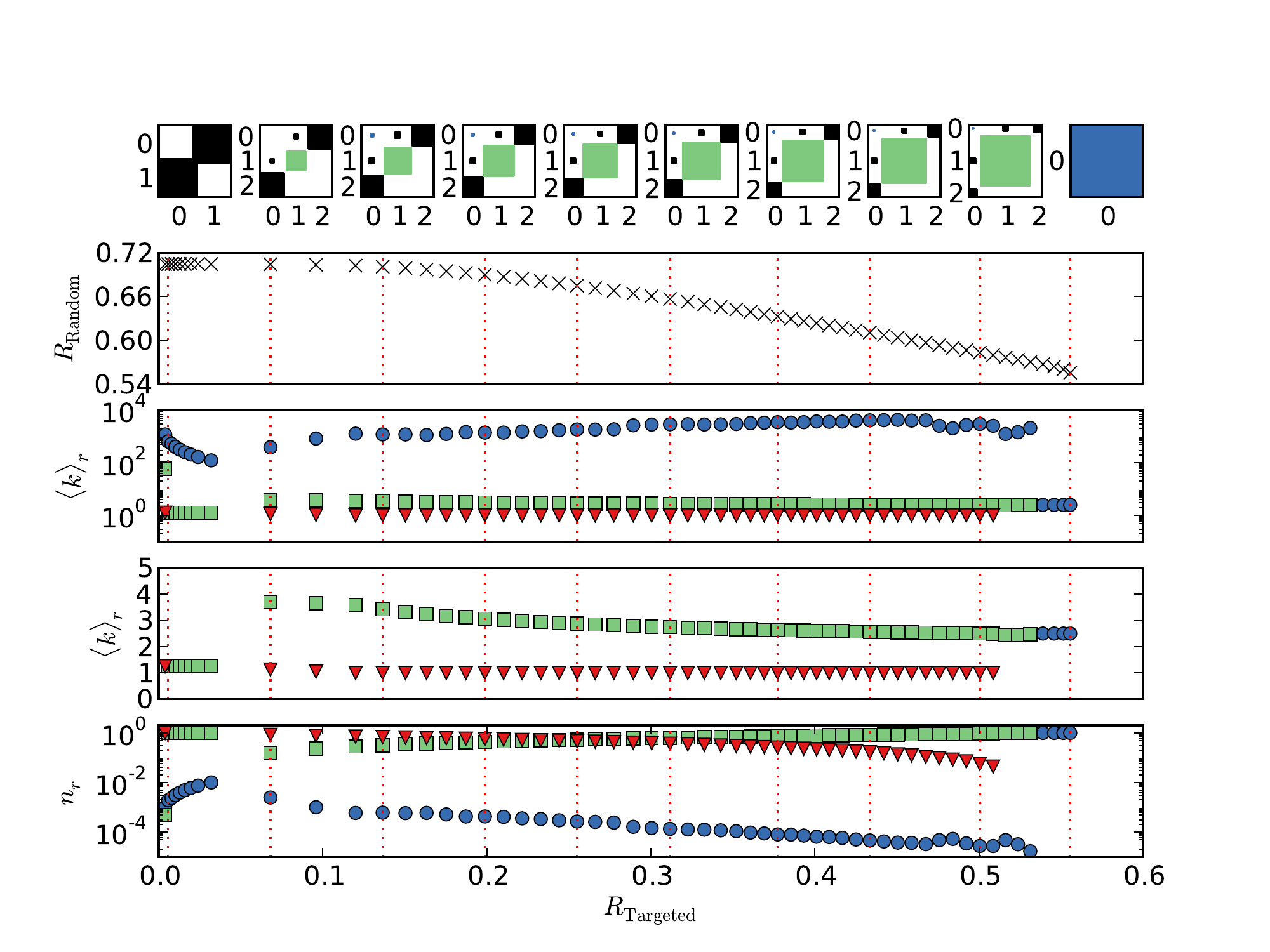} 
  \caption{{\bf Parameters of the optimized networks as a function of \ensuremath{R_\mathrm{Targeted}}
    obtained from a three-block optimization with $\km = 2.5$.}
    The upper row shows the elements of the edge matrix $e_{rs}$,
    where the areas of the squares is proportional to the logarithm of the element. 
    The positions for which these Hinton plots are shown are marked with
    dashed lines in the other panels. The second row shows the trade-off curve
    already displayed in Fig.~\ref{fig_pareto_2blocks},
    while the third and fourth display the mean degree of the blocks, 
    on a logarithmic and a linear scale, respectively.
    The last row shows the relative sizes of the blocks.
    The coloring  of the blocks and their index is determined by their mean degree,
    where the block with the highest mean degree is shown in blue and always has
    index $r=0$, followed by green with $r=1$ (second highest) and red
    ($r=2$,  lowest degree).
    \label{fig_k2.5_3Blocks}}  
\end{figure}

\begin{figure}[th!]
   \centering
   \includegraphics[angle=0, width=\columnwidth]{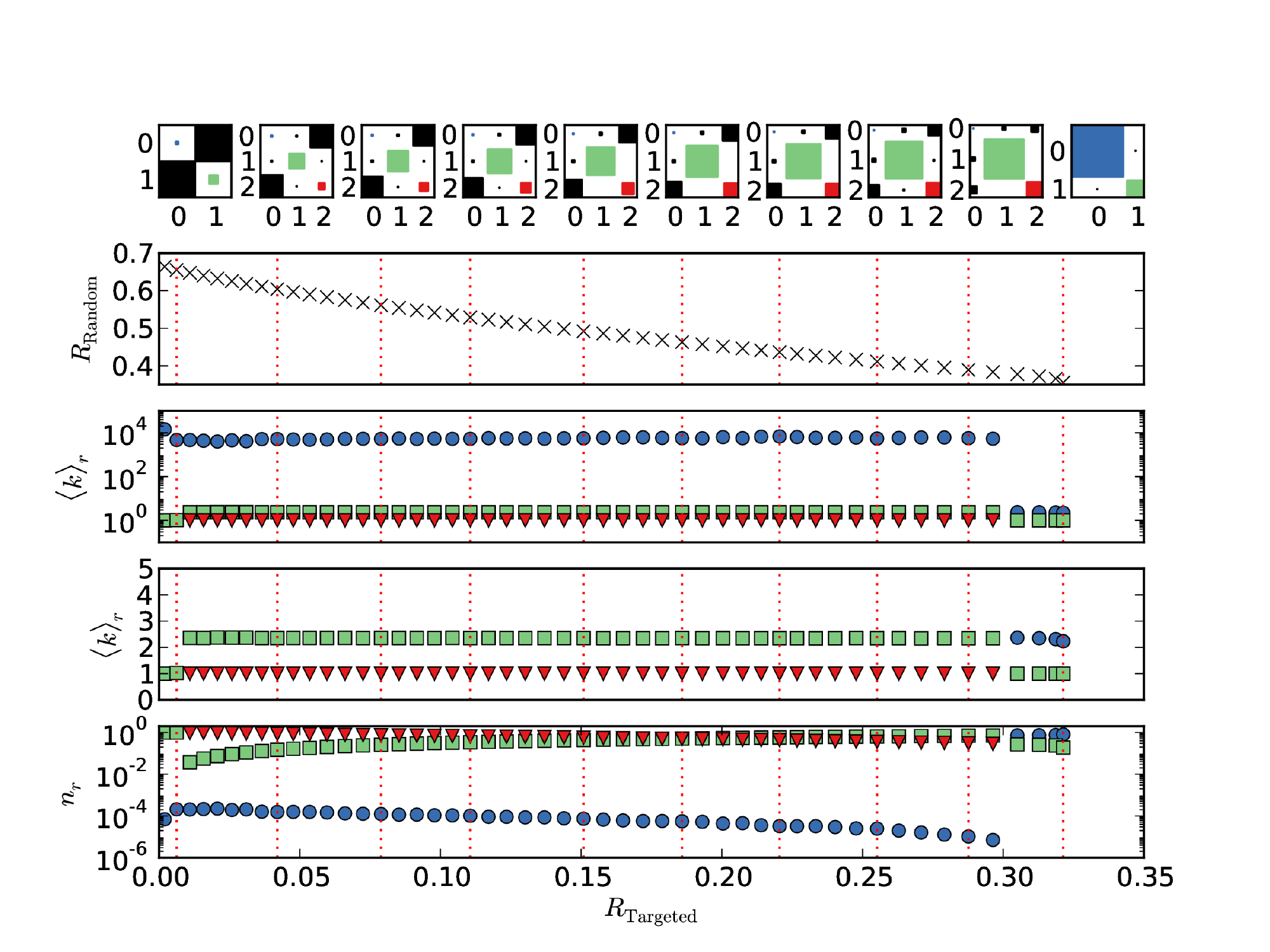} 
   \caption{{\bf Parameters of the networks along the
     trade-off curve for the five block optimization with $\km =
     2.0$.} The rows show the elements of the edge matrix $e_{rs}$ ,
the trade-off curve already displayed in Fig.~\ref{fig_pareto_2blocks}, the mean degree of
the blocks on a logarithmic and on a linear scale, as well as the relative sizes of the
blocks, from top to bottom, respectively. See caption of Fig.~\ref{fig_k2.5_3Blocks} for more details on the
coloring and box sizes.
     \label{fig_k2.0_3Blocks}
   }
\end{figure}

\begin{figure}[th!]
   \centering
   \includegraphics[width=\columnwidth]{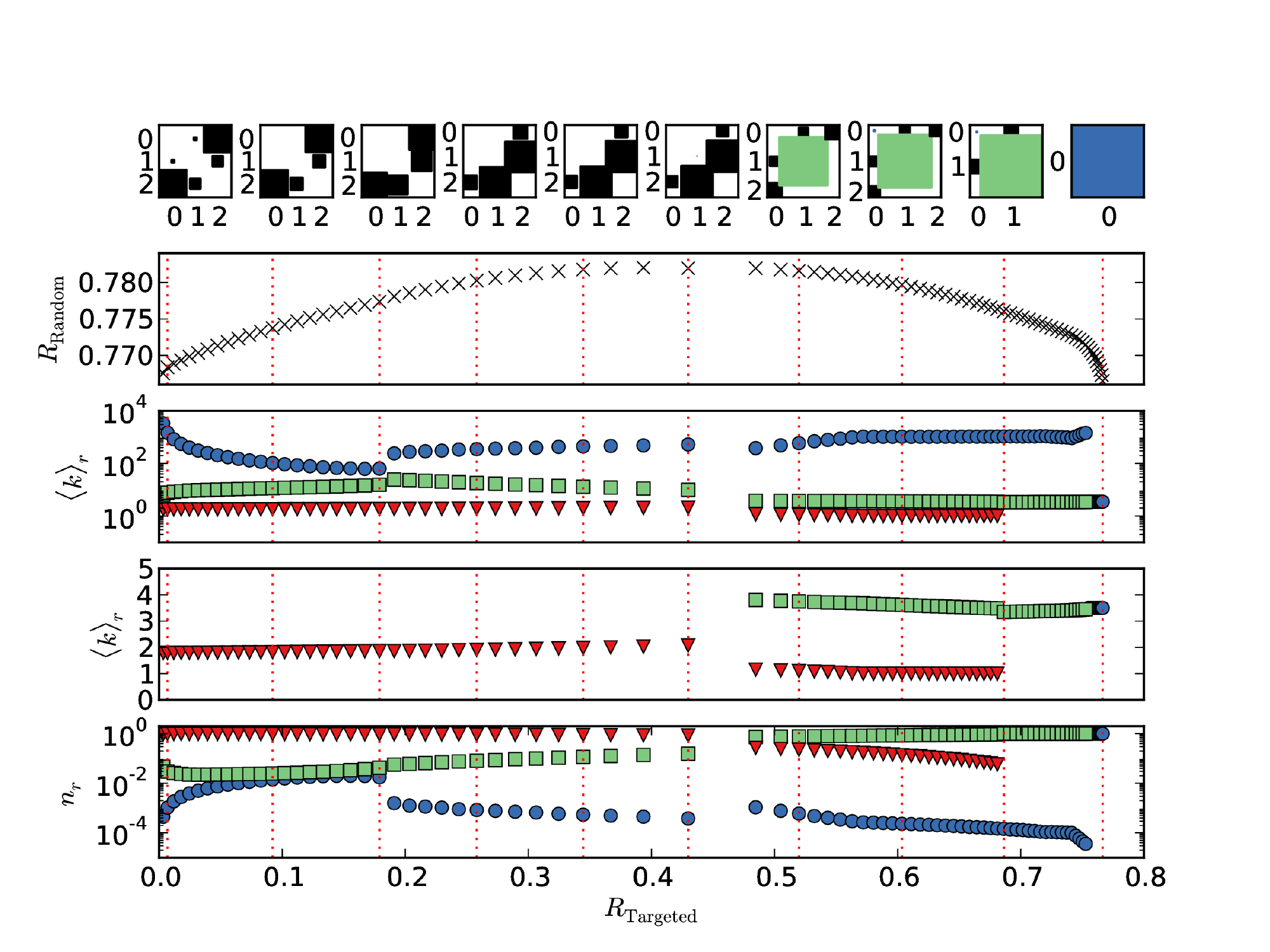} 
   \caption{{\bf Parameters of the networks along the
     trade-off curve for the five block optimization with $\km =
     3.5$.} The rows show the elements of the edge matrix $e_{rs}$ ,
the trade-off curve already displayed in Fig.~\ref{fig_pareto_2blocks}, the mean degree of
the blocks on a logarithmic and on a linear scale, as well as the relative sizes of the
blocks, from top to bottom, respectively. See caption of Fig.~\ref{fig_k2.5_3Blocks} for more details on the
coloring and box sizes.
     \label{fig_k3.5_3Blocks}}  
\end{figure}

\section*{Acknowledgments}

The authors acknowledge fruitful discussions with Barbara Drossel.
C.P.\ and S.S.\ additionally acknowledge the support
and discussions with colleagues at the Honda Research Institute
Europe GmbH.
T.P.P.\ was funded by the University of Bremen, zentrale
Forschunsf\"orderung Linie 04. C.P.\ was funded by the Honda Research Institute Europe GmbH.

\bibliography{robustness.bib}

\end{document}